\documentstyle[12pt]{article}
\input epsf
\parskip=\medskipamount
   \def\CL{{\cal L}}

   \def\Ga{\Gamma}

\def\la{\lambda}  
\def\te{\theta}   
\def\om{\omega}   \def\Om{\Omega}
   
\def\IC{\relax{\rm l\kern-.50 em C}}
\def\IE{\relax{\rm l\kern-.12 em E}}
\def\IK{\relax{\rm l\kern-.18 em K}}
\def\IL{\relax{\rm I\kern-.18 em L}}
\def\IN{\relax{\rm I\kern-.18 em N}}
\def\IR{\relax{\rm I\kern-.18 em R}}
\def\ii{\rm i\,}
\font\tenfrak=eufm10  \font\sevenfrak=eufm7  \font\fivefrak=eufm5
\newfam\frakfam
\textfont\frakfam=\tenfrak
\scriptfont\frakfam=\sevenfrak
\scriptscriptfont\frakfam=\fivefrak

\newtheorem{proposicion}{Proposition}

\def\wt{\widetilde}
\def\frac#1#2{{#1\over #2}}
\def\fracpd#1#2{\frac{\partial #1}{\partial #2}}

\begin{document}

\title{
  Lagrangian Formalism for nonlinear second-order Riccati Systems:
  one-dimensional Integrability and two-dimensional Superintegrability }

\author{
Jos\'e F. Cari\~nena$\dagger\,^{a)}$,
Manuel F. Ra\~nada$\dagger\,^{b)}$,
Mariano Santander$\ddagger\,^{c)}$ \\
$\dagger$
  {\sl Departamento de F\'{\i}sica Te\'orica, Facultad de Ciencias} \\
  {\sl Universidad de Zaragoza, 50009 Zaragoza, Spain}  \\
$\ddagger$
  {\sl Departamento de F\'{\i}sica Te\'orica, Facultad de Ciencias} \\
  {\sl Universidad de Valladolid,  47011 Valladolid, Spain}
}
\maketitle
\date{}

\begin{abstract}
The existence of a Lagrangian description for the second-order
Riccati equation is analyzed and the results are applied to the
study of two different nonlinear systems both related with the
generalized Riccati equation.
The Lagrangians are nonnatural and the forces are not derivable
from a potential.
The constant value $E$ of a preserved energy function can be used
as an appropriate parameter for characterizing the behaviour of the
solutions of these two systems.
In the second part the existence of two--dimensional
versions endowed with superintegrability is proved.
The explicit expressions of the additional integrals are obtained in 
both cases.
Finally it is proved that the orbits of the second system, that represents
a nonlinear oscillator, can be considered as nonlinear Lissajous figures
\end{abstract}

\begin{quote}
Nonlinear oscillations. Closed trajectories.

{\sl Running title:}{\enskip}
Lagrangian Formalism for Riccati Systems.



\end{quote}
{\vfill}

\footnoterule
{\noindent\small
$^{a)}${\it E-mail address:} {jfc@unizar.es}  \\
$^{b)}${\it E-mail address:} {mfran@unizar.es} \\
$^{c)}${\it E-mail address:} {santander@fta.uva.es}
\newpage

\section{Introduction }

Ince studied, in his well-known book of differential equations \cite{In56},
the following equation
$$
  w'' + 3 w w' + w^3 = q(z)
$$
and proved that it has the general solution $w=u'/u$, where $u$ is
a general solution of the linear equation of the third order
$u'''=q(z)u$. This equation was also studied by Davis in
\cite{Da62} as a particular case of the generalized Riccati
equations (according to Davis the family of these nonlinear
equations was first studied by E. Vessiot in 1895 and G.
Vallenberg in 1899; see \cite{Er76,Er77,GrL99} for some more
recent studies related with higher-order Riccati equations). Later
on Leach {\sl et al} \cite{Le85,LeF88} consider the equation
\begin{equation}
  \ddot{q} + q \dot{q} + \beta\,q^3  =  0
\label{EqLeach}\end{equation}
and point out that ``for $\beta =
1/9$ is linearizable, possesses eight symmetries and is completely
integrable" and they add ``consequently, we could expect that this
remarkable mathematical property corresponds to an important
physical one appearing (or disappearing) for this value which
consequently would appear as a critical one". This particular
$\beta = 1/9$ equation was also obtained in \cite{DD87} in the
study of nonlinear equations with the maximum number of symmetries
(see \cite{MaL85}, \cite{SaM87}, \cite{MaL89}, and \cite{KaM94}
for the Lie symmetry approach to dynamical systems).

Recently Chandrasekar {\sl et al} \cite{ChS04} have studied a generalization of
this equation obtained as a particular case of the Lienard  equation
$$
  \ddot{x} + f(x) \dot{x} + g(x)  =  0
$$
given by $f(x) = k\,x$ and $g(x) = (1/9)k\,x^3 + {\lambda} x$.
Although this new equation also belongs to the generalized Riccati family
studied by Davis and Leach {\sl et al}, they make use of a two step
procedure to solve the problem:
firstly they use the so-called Prelle-Singer method
\cite{PrS83,DD01,ChS04b,ChS04c}
for obtaining a set of time-dependent integrals of motion and
secondly they use this time-dependent family in order to compute
the solution.
The result is interpreted, when $\la>0$, as an
``unusual Li\'enard type oscillator with properties of a linear 
harmonic oscillator".
But we wish to call the attention to one property discussed in the final
part of the paper (after finalizing with the Prelle-Singer method):
the existence of a Lagrangian description.

  The main objective of this article is to develop a deeper analysis
of these nonlinear  equations using the Lagrangian formalism as an
approach.
In fact the starting point of our approach is the fact that the
Riccati equation belongs to a family of nonlinear equations admitting
a Lagrangian description.
This has interesting consequences, the most important of them
is that Riccati systems are systems endowed with a preserved energy function.
We study the two nonlinear systems firstly in one dimension and
then in two dimensions.
Moreover we prove that the two-dimensional extensions are not only
integrable but also super-integrable.
We note that this situation has certain similarity with the one-dimensional
nonlinear oscillator studied by Mathews {\sl et al} in  \cite{MaL74}
that has been proved to admit a superintegrable two-dimensional
version \cite{CaR04}.

  The plan of the article is as follows:
In Sec. 2 we present a Lagrangian approach to a family of nonlinear equations
that includes the second-order Riccati equation as a particular case.
Sec. 3, that is devoted to the first nonlinear system
(`dissipative'-looking system), is divided in three parts
corresponding to the one-dimensional system, geometric
formalism and symmetries, and two-dimensional system
and super-integrability, respectively.
Sec. 4, that is devoted to the second nonlinear system (`nonlinear
oscillator'), also firstly studies the $n=1$ system and then the
two-dimensional system that, as we have pointed out, is also endowed
with super-integrability.
Finally in Sec. 5 we make some comments.

\section{Lagrangian formalism and second-order Riccati equations}

In this article we shall consider the following nonlinear
second-order equation
\begin{equation}
  y'' + \bigl[\,b_0(t) + b_1(t) y\,\bigr] y' +
      a_0(t) + a_1(t) y + a_2(t) y^2 + a_3(t) y^3 = 0 \,,
\label{Riccati2}\end{equation}
where we suppose that $a_3>0$ and
the two functions $b_0$, $b_1$, are not independent but satisfy
$$
b_0 = \frac{a_2}{\sqrt{a_3}} - \frac{a_3'}{2 a_3} \,,\quad
b_1 = 3 \sqrt{a_3} \,.
$$
The more important property of this equation is that it can be
transformed into a third-order linear equation by the substitution
$$
  y(t) = \frac{1}{\sqrt{a_3(t)}}\,\frac{v'(t)}{v(t)} \,.
$$
Thus the equation (\ref{Riccati2}), that is the natural
second-order generalization of the well known Riccati equation,
  is therefore a nonlinear equation the solution of
which can be expressed in terms of solutions of a linear equation
of third order. In fact it can be considered as the particular
$n=2$ case of a more general situation that can be approached by
Lie theory or by the action of an operator $R$.  From the Lie
theory of symmetries of differential equations, the invariance of
the $n$-order linear equation,
$$
  v^{(n)} + p_1(t) v^{(n-1)} + \dots + p_n(t)v = 0 \,,
$$
under the vector field $X= v\,({\partial}/{\partial}v)$, that
represents the infinitesimal generator of dilations, means the
existence of a change $v=e^u$ such that $X$ becomes
$X={\partial}/{\partial}u$, and the transformed equation reduces
to an equation of order $(n-1)$ for $y=u'$ that for the $n=2$
reduces to the usual first-order Riccati equation.
  Alternatively if $R$ denotes the following differential operator
$$
  R = \frac{d}{dt} + y(t)\,,
$$
then the Riccati equation of order $n$ is given by
$$
  (R^n + p_1 R^{n-1} + \dots + p_{n+1} R +  p_n)y + p_{n+1} = 0
$$
where $p_j=p_j(t)$, $j=1,2,\dots,n+1$, are $n+1$ arbitrary
functions. We note that these two methods lead to the subfamily of 
the nonlinear
equations with the coefficient of the higher power equal to one;
nevertheless the general Riccati equation is generated by the change 
$t=f(\tau)$
of the independent variable; for example for $n=2$ we obtain
$$
  y'' + \bigl[\,p_1 + 3 y\,\bigr] y' + p_3 + p_2 y + p_1 y^2 + y^3 = 0
$$
and after the time reparametrization we arrive to
$$
  \frac{d^2y}{d\tau^2} + \Bigl[\,(f'p_1 - \frac{f''}{f'}) + 3 f' 
y\,\Bigr] \frac{dy}{d\tau}
  + (f'^2p_3) + (f'^2p_2)\, y + (f'^2p_1)\, y^2 + f'^2\, y^3 = 0
  \,.
$$

  We are interested in the study of nonlinear systems given by (\ref{Riccati2})
but first, in this section, we consider a more general familly from
which the second-order Riccati equation appears as a particular case.

  At this point we recall that a Lagrangian function $L$ is called `natural'
or `of mechanical type' when it is of the form $L=T-V$, where $T$ is a
quadratic  kinetic term and $V$ is a potential function.
Most of the known Lagrangian equations arise from Lagrangians of this
particular type;
nevertheless the Lagrangian formalism is well defined, not only for these
specific functions but also for more general Lagrangian functions.

\begin{proposicion}
The nonlinear second-order Riccati equation admits a Lagrangian
description.
\end{proposicion}
{\it Proof:}
  We first consider the following one degree of freedom Lagrangian
\begin{equation}
  L =  \frac{1}{v_x + k\,U(x,t)} \,.   \label{LUt}
\end{equation}
Then we arrive to the following second-order nonlinear equation
\begin{equation}
\frac{d^2 x}{dt^2} + \hbox {$(\frac{3}{2})$} k U_x' \Bigl(\frac{d
x}{dt}\Bigr) +  \hbox {$(\frac{1}{2})$} k^2 U U_x' + k U_t'= 0 \,.
\label{EqU(x,t)}\end{equation}
In the particular case of the
function $U=U(x,t)$ being a quadratic function
$$
  U = c_0(t) + c_1(t) x + c_2(t) x^2 \,,
$$
the above equation (\ref{EqU(x,t)}) reduces to
\begin{equation}
\frac{d^2 x}{dt^2} + \bigl(\,b_0 + b_1 x\,\bigr) \Bigl(\frac{d
x}{dt}\Bigr) + a_0 + a_1 x + a_2 x^2 + a_3 x^3 = 0
\label{Riccati2xt}\end{equation} where the four functions $a_0$,
$a_1$, $a_2$, $a_3$, are given by
\begin{eqnarray}
  a_0 &=& (\frac{1}{2}) c_0 c_1 + c_0'   \,,\quad
  a_1 = c_0 c_2 + (\frac{1}{2}) c_1^2 + c_1'   \,,   \cr
  a_2 &=& \bigl(\frac{3}{2}\bigr) c_1 c_2  + c_2' \,,\quad
  a_3 = c_2^2    \,,
{\nonumber}\end{eqnarray}
and the two functions $b_0$ and $b_1$ satisfy the appropriate restrictions
$$
  b_0  = \bigl(\frac{3}{2}\bigr) c_1\,,\quad   b_1 = 3c_2  \,.
$$
Thus the second-order Riccati equation (\ref{Riccati2xt}), that is
a particular case of the equation (\ref{EqU(x,t)}), is the
Euler-Lagrange equation of the Lagrangian function (\ref{LUt})
in the particular case of a quadratic function $U$.

  As a corollary of this proposition we can state that, if the function $U$
is time-independent, then the nonlinear equation (\ref{EqU(x,t)}) has a
first-integral that can be interpreted as a preserved energy.
The idea is as follows: if we restrict our study to the case of
time-independent systems, that is, to nonlinear equations arising
from a Lagrangian of the form
\begin{equation}
  L =  \frac{1}{v_x + k\,U(x)}     \label{LU}
\end{equation}
then we can define an associated Lagrangian energy $E_L$ by the
usual procedure
$$
  E_L = \Delta(L) - L \,,\quad   \Delta = v_x\,\fracpd{}{v_x} \,,
$$
and we arrive to
$$
  E_L = \frac{-\,\,(2\,v_x +\,k\,U(x))}{(v_x + k\,U(x))^2}
\,,\qquad \frac{d}{dt}E_L = 0.
$$

  Note that $L$ is non-natural and, as there is neither kinetic term
$T$ nor potential function $V$, the energy can not be of the form
$E_L=T+V$.
Note also that a Lagrangian is defined up to certain ambiguities;
that is, $\wt{L} = c_1 L + c_0$ determines the same differential
equation but it leads to an slightly different energy $\wt{E_L}$
given by $\wt{E_L} = c_1 E_L - c_0$. In the `natural' case, $c_1$ is
determined by the corresponding Riemannian metric (hence the
classical one half coefficient) and $c_0$ is absorbed in the
potential; here we have just taken $c_1=1$, $c_0=0$. Concerning
the negative sign, that could be considered as something
inconvenient, it does not matter at all (it is only an aesthetic
question); in fact it can be removed just by chosen $\wt{L} = - L$
as a new Lagrangian.

   As for the `natural' case we can obtain the solution of the dynamics
from the conservation law of the energy.
If we assume that $E_L$ takes the constant value $E_L= E$, then we arrive to
$$
  E\,v_x^2 + 2\bigl(1 + k\,E\,U(x)\bigr) v_x + k\,\bigl(1 + 
k\,E\,U(x)\bigr) U(x) =0
$$
and on solving for $v_x$ and making separation of variables we arrive at
\begin{equation}
  t - t_0= \ -\  \int_{x_0}^x {\frac{E\ dx}{(1 + k\,E\,U(x)) \,\pm\,\sqrt{\,1
  + k\,E\,U(x)}}}  \,.
\label{soleqU}\end{equation}
The motion is confined to the region where $E{\ge}-1/(k U)$.
To sum up, a time-independent system describedby the nonlinear equation
(\ref{EqU(x,t)}) is solvable and the solution of the dynamics
is given (up to one integration) by (\ref{soleqU}).

  An remarkable property is that the conservation of the energy $E_L$
leads to a plus/minus sign in the expression for the velocity
$$
  \frac{dx}{dt} = \frac{-\,(1 + k\,E\,U(x)) \,\pm\,\sqrt{\,1+ 
k\,E\,U(x)}}{E}  \,.
$$
Thus, we obtain two different values for the velocity at the same
point $x$. It is known that in the standard case of a particle in
a potential $V(x)$ we also have two possibilities but both with
same modulus (the positive value for the motion from left to right
and the negative for the opposite motion from right to left).
In this case the situation is different; we have a Lagrangian but
not a potential and the two
possible values differ, not just in the sign, but in the absolute
value; thus the motions from left to right and from right to left
take place at different velocities.

We close this section with the problem of the existence of
alternative Lagrangians.

In differential geometric terms a time-independent Lagrangian
function $L$ determines an exact two-form $\om_L$ defined as
$$
  \te_L = \Bigl(\fracpd{L}{v_x}\Bigr) \,dx  \,,\qquad
  \om_L = -\,d\te_L  \,,
$$
in such a way that, when $L$ is nonsingular, $\om_L$ is symplectic 
and the dynamics is
given by the vector field $\Ga_L$ solution of the equation
$$
  i(\Ga_L)\,\om_L = dE_L   \,.
$$
In this particular case $\om_L$ and $\Ga_L$ are given by
$$
  \om_L = \frac{2\,dx{\wedge}dv_x}{(v_x + k\,U(x))^3} \,,\qquad
  \Ga_L = v_x\,\fracpd{}{x} + F_x\,\fracpd{}{v_x}   \,,\qquad
  F_x = -\,(\frac{1}{2})\,k\,\bigl(3 v_x + k U(x)\bigr) U_x' \,.
$$
An important property of the Lagrangian formalism is that for one
degree of freedom systems there exist many different equivalent
Lagrangians \cite{CuS66,HoH81}. A sketch of the proof is as
follows: in a two-dimensional manifold all the symplectic forms
must be proportional. Hence, for a one degree of freedom
Lagrangian, any other symplectic form $\om_2$ must be proportional
to $\om_L$, that is $\om_2 = f\om_L$. Then
$$
  i(\Ga_L)\,\om_2 = f\,i(\Ga_L)\,\om_L = f\,dE_L
$$
The right-hand side is an exact one-form if, and only if, $df{\wedge}dE_L=0$,
which shows that $f$ must be a function of $E_L$.
In this case it can be proved that the new symplectic form $\om_2$ is derivable
from an alternative Lagragian $L_2{\ne}L$ for $\Ga_L$.

  In this particular case, starting with the Lagrangian (\ref{LU}) and
assuming for the constant of motion $f$ the particular expression $f= 
(-1/E_L)^{3/2}$,
we have obtained the following function
\begin{equation}
  L_2 = \sqrt{2\,v_x +\,k\,U(x)}   \label{Lkn1b}
\end{equation}
as a new alternative Lagrangian for the $t$-independent version of the
equation (\ref{EqU(x,t)}).
This new Lagrangian, that is neither `natural' or `of mechanical type',
is equivalent to $L$ in the sense that both determine the same dynamics.
Nevertheless we must say that it is not clear whether $L_2$ will lead to
simpler expressions for other dynamical properties;
so, in the following, we only use the original Lagrangian (\ref{LU}).

\section{Lagrangian conservative approach to a `dissipative'-looking
nonlinear system}

\subsection{One-dimensional nonlinear system: energy and integrability}

  We now apply the formalism introduced in Sec. 2 to the study of
the following nonlinear equation
\begin{equation}
\frac{d^2 x}{dt^2} + 3 k x \Bigl(\frac{d x}{dt}\Bigr) + k^2 x^3  =
0  \,.
\label{Eqkn1}\end{equation}
It is a special case of
equation (\ref{Riccati2}) and because of this is a Lagrangian
equation. In fact, it is easy to verify that it can be obtained
from the following Lagrangian function
\begin{equation}
  L =  \frac{1}{v_x + k\,x^2}  \,. \label{Lkn1}
\end{equation}
Two important properties are:

(i) Firstly we note that the value $\beta=1/9$, pointed out by
Leach {\sl et al} \cite{LeF88} as the particular value introducing
a high degree of regularity in the nonlinear problem
(\ref{EqLeach}), appears now as related with the Lagrangian origin
of the equation. That is, only if $\beta=1/9$ the equation
(\ref{EqLeach}) belongs to the Lagrangian family (\ref{Eqkn1})
arising from (\ref{Lkn1}).

  (ii) Secondly this equation looks like a dissipative equation
  with the damping term proportional to $x v_x$; nevertheless
  it is in fact a conservative system because of its Lagrangian origin.
  What happens is that the term conservative is usually considered in the
Newtonian sense, that is, a particle moving in a one-dimensional
potential and forces determined as the gradient of the potential.
Here the force is a velocity-dependent force and conservative just
means the existence of a preserved (non-Newtonian) energy function
that is given by
$$
  E_L = \frac{-\,\,(2 v_x + k\,x^2)}{(v_x + k\,x^2)^2} \,.
$$

  Next we turn to the solution of the dynamics. Instead of considering
directly the nonlinear equation we can solve this problem by making
use of the conservation law of the energy;
if we assume $E_L = E$, then we arrive at
$$
  E\,v_x^2 + 2 (1 + k\,Ex^2)\, v_x + k\,(1 + k\,Ex^2)\,x^2 =0
$$
and solving for $v_x$ we obtain
$$
  \frac{dx}{dt}= \frac{-\,(1 + k\,Ex^2) \pm \sqrt{\,1 + k\,Ex^2}}{E}  \,.
$$
So, after integration, we arrive at
$$
  t = \frac{1}{kx}\,\bigl(1 \pm \sqrt{1 + k\,Ex^2} \,\bigr)
$$
which yields
$$
  x = \frac{2 t}{k t^2 - E}
$$
that represents the solution of the dynamics as a function of the
constant value $E$ of the Lagrangian energy (for easy of notation we
give the solution for the particular initial conditions $(t_0=0,x_0=0)$).
The system is well defined for $(k>0,E<0)$ and $(k<0,E>0)$, but
for $(k<0,E<0)$ or $(k>0,E>0)$ it is singular at
$t={\pm}\sqrt{|E|/|k|}$.
In fact the double change $(k,E)\to(-k,-E)$ is equivalent to a
time-inversion (see Figure I).

  Since $x(t)$ is the quotient of two polynomials in $t$, with the
denominator of higher degree than the numerator, the trajectories
approach the origin when $t$ increases. How can this behaviour be
compatible with the conservation of the energy? An analysis of the
expression that $E_L$ takes on the trajectories shows that it
reduces to the quotient of two functions both going down as
$t{\to}\infty$, but in such a way that the ratio remains
constant. Moreover it can also be proved that the velocity $v_x$
decreases as $t{\to}\infty$ in such a way that the particle
approaches but never reaches the origin in the phase plane. The
important point is that the dependence of $E_L$ with respect to
$v_x$ is defined in such a way that even when $dx/dt$ decreases
the value of $E_L$ remains constant.

  The phase space analysis shows that the origin is a nonelementary
critical point for which the linear approximation is not valid.
If we consider a small neighbourhood of the point, then we find, when
$k>0$, that it has four different sectors \cite{Pe01} (see Figure
II): an elliptic sector (upper side), an hyperbolic sector (lower
side) and two parabolic sectors (a source on the left and a sink
on the right).

\subsection{Symplectic formalism and master symmetries}

  It is known that the nonlinear equation (\ref{Eqkn1}) admits constants
of motion depending explicitly on the time $t$
\cite{ChS04},\cite{ChS04b}. Now we  prove that this property
is a related with the fact that the Lagrangian (\ref{Lkn1}) admits
``master symmetries". This is in fact an important property, not
only from the geometric point of view but also because it is
directly related with the superintegrability of the
two-dimensional version of this system.

  A function $T$ that satisfies the following property
$$
  {d \over dt}\,T \ne  0  {\quad},\dots,{\quad} {d^m \over dt^m}\,T \ne 0
  \,,\quad {d^{m+1} \over dt^{m+1}}\,T=0 \,,
$$
is called a generator of integrals of motion of degree $m$. Notice
that this means that the function $T$ is a non-constant function
generating a constant of motion by time derivation. If we denote
by  $T_{x1}$ and $T_{x2}$, the functions
$$
  T_{x1} = \frac{1}{v_x + k\,x^2}  \,,\qquad
  T_{x2} = \frac{x}{v_x + k\,x^2}  \,,
$$
then we have
$$
  {d \over dt}\,T_{x2} = 1  \,,{\quad} {d \over dt}\,T_{x1} = 
k\,T_{x2}  \,,{\quad}
  {d^2 \over dt^2}\,T_{x1} = k  \,,{\quad}   {d^3 \over dt^3}\,T_{x1} = 0  \,.
$$
Hence $T_{x1}$ and $T_{x2}$ are generators of integrals of motion for
the Lagrangian (\ref{Lkn1}).

  In geometric terms this property is related with the existence
of ``master symmetries" \cite{SaC81,Da93,Fe93,Ra99,Da04}.
If the dynamics is represented by a certain vector field $\Ga$;
then a vector field $Z$ that satisfies the following two properties
$$
  [Z,\Ga] \,=\,{\widetilde Z} \ne 0  \,,{\qquad}
  [\,{\widetilde Z}\,,\Ga]  = 0  \,,
$$
is called a ``master symmetry" of degree $m=1$ for $\Ga$.
If $Z$ is such that
$$
  [Z,\Ga] = {\widetilde Z} \ne 0  \,,\
  [\,{\widetilde Z}\,,\Ga]   \ne 0  \quad {\rm and}\quad
  [\,[\,{\widetilde Z}\,,\Ga]\,,\Ga] = 0 \,,
$$
then it is called a ``master symmetry" of degree $m=2$.

  We focus our attention in the case of master symmetries
given rise via ${\widetilde Z}$ to constants of motion. Let $L$ be
a time-independent Lagrangian,  $Z$ the Hamiltonian vector field
of a certain function $T$ and suppose that $Z$ is a
(time-independent) master symmetry of  $\Ga_L$. Then the
time-dependent vector field $Y_Z$ determined by $Z$ and defined as
$$
   Y_Z = Z + t\,[Z,\Ga_L] +
   (\frac{1}{2})\,t^2\,[\,[Z,\Ga_L]\,,\Ga_L]+ \dots
$$
is a time-dependent symmetry of $\Ga_L$ that, in the case $m=1$,
satisfies
$$
  i(Y_Z)\,\Om_E = d\,[\, T - t\,\Ga_L(T)\,]
$$
where $\Om_E = \om_L + dE_L{\wedge}dt$. Hence the time-dependent
function $J^t = T - t\,\Ga_L(T)$ is a time-dependent constant of
motion (for $m=2$ the corresponding constant $J^t$ is quadratic in
$t$)

  We now return to the Lagrangian (\ref{Lkn1}) and denote
by $Z_{x1}$ and $Z_{x2}$, the Hamiltonian vector fields of
$T_{x1}$ and $T_{x2}$,
$$
  i(Z_{x1})\,\om_L = dT_{x1}  \,,\qquad
  i(Z_{x2})\,\om_L = dT_{x2}  \,,
$$
which are given by
\begin{eqnarray}
  Z_{x1} &=& -\,(\frac{1}{2}) M_x\,
\Bigl(\, \fracpd{}{x} - 2 k x\,\fracpd{}{v_x}\,\Bigr)  \,,\cr
  Z_{x2} &=& -\,(\frac{1}{2}) M_x\,
\Bigl(\, x\,\fracpd{}{x} + (v_x - kx^2)\,\fracpd{}{v_x}\,\Bigr) \,,
{\nonumber}\end{eqnarray}
where $M_x$ denotes $M_x=v_x + k\,x^2$.
Then we have
$$
  [Z_{x1}\,,\,\Ga_L] = - k\,Z_{x2}  \,,\qquad
  [\,[\,Z_{x1}\,,\Ga_L]\,,\Ga_L] =0 \,.
$$
It is clear that $Z_{x1}$ is a master symmetry. Concerning
$Z_{x2}$ it has some very interesting characteristics; it is a
dynamical symmetry (that is, $[ Z_{x2}\,,\,\Ga_L] = 0$) and it is
a symplectic symmetry (that is, ${\CL}_{Z_{x2}}\,\om_L = 0$), but
nevertheless it is not a Cartan symmetry of the Lagrangian system
$\Ga_L$ because $Z_{x2}(E_L){\ne}0$.
Recall that the two properties ${\cal L}_X\om_L=0$ and $[X\,,\,\Ga_L] = 0$
imply that ${\cal L}_XdE_L=0$, but from this we only obtain that
$X(E_L)$ must be a numerical constant.

  As explained above, the vector fields  $Z_{x1}$ and $Z_{x2}$, determine
two new vector fields $Y_1$ and $Y_2$, given by
\begin{eqnarray}
  Y_1 &=& Z_{x2} + t\,[Z_{x2},\Ga_L] = Z_{x2}  \,,\cr
  Y_2 &=& Z_{x1} + t\,[Z_{x1},\Ga_L] + 
(\frac{1}{2})\,t^2\,[\,[Z_{x1},\Ga_L]\,,\Ga_L]
       = Z_{x1} - k\,t\,Z_{x2}  \,,
{\nonumber}\end{eqnarray}
such that they satisfy
\begin{eqnarray}
  i(Y_1)\,\Om_E &=& i(Z_{x2})\,\om_L + Z_{x2}(E_L)\,dt = d\,[\, 
T_{x2}\,-\,t\,] \,,\cr
  i(Y_2)\,\Om_E &=& i(Z_{x1} - k\,t\,Z_{x2})\,\om_L + Z_{x1}(E_L)\,dt
             - k\,t\,Z_{x2}(E_L)\,dt
= d\,[\, T_{x1} - k\,t\,T_{x2} + (\frac{1}{2}) k\,t^2   \,]    \,.
{\nonumber}\end{eqnarray}
Hence the two following functions
$$
  J_{x1}^t = T_{x2} - t   \,,\quad
  J_{x2}^t = T_{x1} - k\,t\,T_{x2} + (\frac{1}{2}) k\,t^2
$$
are time-dependent integrals of motion determined by
$Z_{x1}$, $Z_{x2}$, via $Y_1$, $Y_2$.

  In the next subsection we see that these symmetries and these time-dependent
integrals are the origin of the $n=2$ superintegrability.

\subsection{Two-dimensional nonlinear system: Lagrangian formalism
and super-integrability}

  We now want to study the nonlinear system
\begin{eqnarray}
  \frac{d^2 x}{dt^2} + 3 k_1 x \bigl(\frac{d x}{dt}\bigr) + k_1^2 x^3 
&=& 0  \cr
  \frac{d^2 y}{dt^2} + 3 k_2 y \bigl(\frac{d y}{dt}\bigr) + k_2^2 y^3  &=& 0
\label{Eqkn2}
\end{eqnarray}
representing the $n=2$ version of the nonlinear equation (\ref{Eqkn1}).
It is clear, from the results of $n=1$, that these two equations
can be considered as the Lagrange equations arising from the
following two-dimensional Lagrangian
\begin{equation}
  L =  \frac{1}{v_x + k_1\,x^2} + \frac{1}{v_y + k_2\,y^2}  \,.
\label{Lkn2}\end{equation}
Therefore the dynamics is characterized by preserving the
following Lagrangian energy
$$
  E_L =  -\,\frac{(2 v_x + k_1\,x^2)}{(v_x + k_1\,x^2)^2} -
  \frac{(2 v_y + k_2\,y^2)}{(v_y + k_2\,y^2)^2}   \,.
$$
The first consequence of the Lagrangian character of the equations
is that, as there is no coupling between the two degrees of freedom,
the two one-dimensional energies are integrals of motion
$$
  I_1 = -\,\frac{(2 v_x + k_1\,x^2)}{(v_x + k_1\,x^2)^2} \,,{\quad}
  I_2 = -\,\frac{(2 v_y + k_2\,y^2)}{(v_y + k_2\,y^2)^2} \,,{\quad}
  \frac{d}{dt}\,I_i = 0  \,,{\quad}  i=1,2.
$$

  Our main goal in the study of this nonlinear problem is to
prove that this system possesses the rather unusual property of
super-integrability. At this point we recall that a system is
called super-integrable if it is integrable in the Liouville-Arnold
sense and, in addition, possesses more independent first integrals
than degrees of freedom
(see Refs. \cite{Ra00}-\cite{RaS03} for some articles published in
these last years and Ref. \cite{Montr04} for a recent workshop on
super-integrability).
It is clear that, for this particular $n=2$ system, super-integrability
means the existence of a third independent integral $I_3$ coupling
the two degrees of freedom in similar way as the angular momentum for
the isotropic linear harmonic oscillator.

The four functions
\begin{eqnarray}
  J_{x1}^t &=& T_{x2} - t     \,,{\quad}
  J_{x2}^t  = T_{x1} - k_1 t\,T_{x2} + (\frac{1}{2}) k_1 t^2   \,,\cr
  J_{y1}^t &=& T_{y2} - t     \,,{\quad}
  J_{y2}^t  = T_{y1} - k_2 t\,T_{y2} + (\frac{1}{2}) k_2 t^2   \,,
{\nonumber}\end{eqnarray}
are time-dependent constants of motion
$$
  \frac{d}{dt}\,J_{xi}^t = 0\,,\qquad  \frac{d}{dt}\,J_{yi}^t = 0 
\,,{\quad}  i=1,2.
$$
We can eliminate the time $t$ by pairing these functions in two different ways
and obtain the following two integrals of motion
\begin{eqnarray}
  I_3 &=&  T_{x2} -  T_{y2}   \,,\cr
  I_4 &=&  k_2 T_{x1} + k_1 T_{y1} -  k_1k_2\,T_{x2}T_{y2}
{\nonumber}\end{eqnarray}
that take the form
\begin{eqnarray}
  I_3 &=& \frac{x}{v_x + k_1\,x^2}  - \,\frac{y}{v_y + k_2\,y^2}   \,,\cr
  I_4 &=& \frac{k_2}{v_x + k_1\,x^2} + \frac{k_1}{v_y + k_2\,y^2}
  - \frac{k_1\,k_2\,x y}{(v_x + k_1\,x^2)(v_y + k_2\,y^2)}     \,.
{\nonumber}\end{eqnarray}
Hence, the nonlinear system given by the equations (\ref{Eqkn2})
and characterizized by the Lagrangian (\ref{Lkn2}) is a superintegrable system.

  In geometric terms the symplectic form $\om_L$ and the dynamical vector field
$\Ga_L$ are given by
\begin{eqnarray}
  \om_L &=& \frac{2\,dx{\wedge}dv_x}{(v_x + k\,x^2)^3}
        + \frac{2\,dy{\wedge}dv_y}{(v_y + k_2\,y^2)^3} \,,\cr
  \Ga_L &=& v_x\,\fracpd{}{x} + v_y\,\fracpd{}{y}
        + F_x\,\fracpd{}{v_x} + F_y\,\fracpd{}{v_y}    \,,
{\nonumber}\end{eqnarray}
where $F_x = -\,k_1x (3 v_x + k_1 x^2)$ and $F_y = -\,k_2 y (3 v_y + k_2 y^2)$.
If we denote by $Z_{xr}$ and $Z_{yr}$, the Hamiltonian vector fields of
$T_{xr}$ and $T_{yr}$, $r=1,2$, then the vector field
$$
  X_3 = Z_{x2} - Z_{y2}
$$
is a dynamical symmetry,
$$
[X_3,\Ga_L]  = 0 \,,
$$
as well a Cartan symmetry,
$$
X_3(E_L)   = 0 \,,\quad \CL_{X_3}\om_L = 0 \,.
$$
It determines the function $I_3$ as the corresponding Hamiltonian
$$
  i(X_3)\,\om_L = dI_3 \,.
$$

   Similarly the vector field $X_4$ given by
$$
  X_4 =  k_2 Z_{x1} + k_1 Z_{y1} - (k_1k_2)\,(T_{y2}Z_{x2} + T_{x2}Z_{y2})
$$
is also a dynamical symmetry as well a Cartan symmetry,
$$
[X_4,\Ga_L]  = 0 \,,\quad
X_4(E_L)   = 0 \,,\quad \CL_{X_4}\om_L = 0
$$
and determines the function $I_4$ as the corresponding Hamiltonian
$$
  i(X_4)\,\om_L = dI_4 \,.
$$

   The main difference between these two symmetries is that $X_3$ is
an exact symmetry of the Lagrangian, that is $X_3(L)=0$, and $X_4$
is a nonexact generalized Noether symmetry
$$
  X_4(L) = \frac{d}{dt}\,F_4 \,,\quad
  F_4 = -\,(\frac{1}{2}) (\frac{k_2}{M_x} + \frac{k_1}{M_y}) \,,
$$
so that $I_4$ is given by
$$
  i(X_4)\te_L - F_4 = I_4 \,.
$$

   An important property related with superintegrability is the
existence of periodic orbits as, for example, in the Kepler
problem or in the harmonic oscillator.
This particular system is superintegrable but the motion,
although bounded, is not  periodic;
instead of this the trajectories are ``almost closed".
We have found that the trajectories in the plane $(x,y)$ are
in fact ``figure eight" curves with the particle making a
complete circuit as $t$ goes from $-\infty$ to $+\infty$.
The trajectory starts very close to the origin, passes through $(0,0)$
only  once and returns once more to the origin, but for $t\to -\infty$
and $t\to \infty$ only approaches $(0,0)$ as a limit.
Figure III shows two curves in the plane $(x,y)$ corresponding
to two different values of $E_1$ and $E_2$.

\section{Lagrangian conservative approach to a nonlinear
oscillator}

\subsection{One-dimensional system: Lagrangian formalism and integrability}

  We now consider the following non-natural Lagrangian
\begin{equation}
  L =  \frac{1}{k v_x + k^2 x^2 + w^2} \,, \label{Lwn1}
\end{equation}
where $k$ and $w$ are arbitrary constants;
the notation $w^2$ for the new parameter clearly advances
that it will be interpreted as a frequency.
We arrive to the following nonlinear equation
\begin{equation}
  \frac{d^2 x}{dt^2} + 3\,k\,x \bigl(\frac{d x}{dt}\bigr) + k^2 x^3 + w^2 x = 0
\label{Eqwn1}
\end{equation}
as well as to the following expression for the Lagrangian energy
$$
  E_L = \frac{-\,\,(2 k v_x + k^2 x^2 + w^2)}{(k v_x + k^2 x^2 + w^2)^2}\,.
$$

We can solve this new nonlinear problem by using the same
technique as in the previous equation (\ref{Eqkn1}). Nevertheless we see
that this new nonlinear equation is in fact a nonlinear oscillator
and, because of this, we will use an approach that is as
close as possible to the linear oscillator.

We denote by $X$ and $W_x$ the following two functions
$$
  X = \frac{x}{(k\,v_x + k^2 x^2 + w^2)}   \,,\quad
  W_x = \frac{v_x + k x^2}{(k\,v_x + k^2 x^2 + w^2)}  \,.
$$
Then we have the following two properties: \\
(i) The time evolution of $X$ and $W_x$ is given by
$$
  \frac{d}{dt}\,X = W_x   \,,{\qquad}
  \frac{d}{dt}\,W_x = -\,w^2\,X   \,.
$$
(ii) $X$ and $W_x$ are related by
$$
  w^2\,X +  k\,x W_x  = x  \,.
$$
Next we see that property (i) is related with the conservation
of the energy and property (ii) with the solution of the dynamics.
In fact, from (i) we conclude that the function $I_{XW}$ defined as
$$
  I_{XW} = W_x^2 + w^2 X^2 \,.
$$
is an integral of the motion.
We remark that, as the system is one-dimensional, the two integrals, 
$E_L$ and $I_{XW}$,
cannot be independent; in fact $I_{XW}$ turns out to be the energy associate to
an equivalent Lagrangian $\wt{L}$ of the form $\wt{L} =  c_1  L + c_2$.
A simple calculation gives
$$
  \wt{E_L} = \Delta(\wt{L}) - \wt{L} = I_{XW}
    \quad {\rm with}\quad
  \wt{L}=  \bigl(\frac{w}{k}\bigr)^2 L - \frac{1}{k^2} \,.
$$
Also from (i) we arrive at
$$
  \frac{d^2}{dt^2}\,X + w^2 X = 0  \,,\qquad
  \frac{d^2}{dt^2}\,W_x + w^2 W_x = 0
$$
so that $X$ and $W_x$ are given by
$$
  X = \bigl(\frac{1}{w}\bigr)\,A\,\sin{(w t + \phi)}  \,,\quad  W_x = 
A\,\cos{(w t + \phi)}
\,,\quad A = \sqrt{E} \,,
$$
where $E$ is the constant value of the new energy function $\wt{E_L}=I_{XW}$.
These two expressions, together with property (ii), lead to the following
trigonometric function for the solution of the dynamics
$$
  x = \frac{w\,\sqrt{E}\,\sin{(w t + \phi)}}{1 - k\,\sqrt{E}\,\cos{(w 
t + \phi)}}\,.
$$

Figure IV represents $x(t)$ as a function of $t$ for several
values of the energy $E$ in the regular oscillatory region
($0<E<1/k^2$). It is clear that for small values of $E$ the
oscillations are rather similar to the oscillations of the linear
system, but for other values of $E$ (roughly speaking, for
$E>0.3/k^2$) the nonlinearity introduces deformations in the
oscillations. For higher values of $E$ the nonlinearity changes
drastically the aspect of the solution. Figure V shows the phase
portrait;
it clearly shows that the motions from left to right and from
right to left take place at different velocities.

\subsection{Two-dimensional system: Superintegrability and nonlinear
Lissajous figures}

We now turn our attention to the two-dimensional version of this nonlinear
oscillator.
If we consider the following Lagrangian
\begin{equation}
  L =  \frac{1}{k_1\,v_x + k_1^2\,x^2 + w_1^2} +
       \frac{1}{k_2\,v_y + k_2^2\,y^2 + w_2^2}   \label{Lwn2}
\end{equation}
then we arrive at the following equations
\begin{eqnarray}
\frac{d^2\,x}{dt^2} + 3\,k_1\,x \bigl(\frac{d x}{dt}\bigr) + k_1^2 
x^3 + w_1^2\,x &=&
0 \,,\cr
\frac{d^2\,y}{dt^2} + 3\,k_2\,y \bigl(\frac{d y}{dt}\bigr) + k_2^2 
y^3 + w_2^2\,y &=&
0   \label{Eqwn2}
\end{eqnarray}
and to the following expression for the energy function
$$
  E_L = -\,\frac{(2\,k_1\,v_x + k_1^2 x^2 + w_1^2)}{(k_1\,v_x + k_1^2 
x^2 + w_1^2)^2} -
  \frac{(2\,k_2\,v_y + k_2^2 y^2 + w_2^2)}{(v_y + k_2 y^2 + w_2^2)^2}   \,.
$$

In this $n=2$ case we have two pairs of functions $(X_j,W_j)$,
$j=1,2$,
\begin{eqnarray}
  X_1 &=& \frac{x}{ k_1\,v_x + k_1^2\,x^2 + w_1^2 } \,,\qquad
  W_1  = \frac{v_x + k_1 x^2}{ k_1\,v_x + k_1^2\,x^2 + w_1^2 } \,,\cr
  X_2 &=& \frac{y}{ k_2\,v_y + k_2^2\,y^2 + w_2^2 } \,,\qquad
  W_2  = \frac{v_y + k_2 y^2}{ k_2\,v_y + k_2^2\,y^2 + w_2^2 }
{\nonumber}\end{eqnarray}
in such a way that we have
$$
  \frac{d}{dt}\,W_j = -\,w_j^2\,X_j   \,,\qquad
  \frac{d}{dt}\,X_j = W_j
$$
and
$$
  \frac{d^2}{dt^2}\,X_j + w_j^2\,X_j = 0  \,,\qquad
  \frac{d^2}{dt^2}\,W_j + w_j^2\,W_j = 0  \,.
$$
A similar calculation shows that
$$
  x = \frac{w_1^2\,X_1}{1 - k_1\,W_1}  \,,{\qquad}
  y = \frac{w_2^2\,X_2}{1 - k_2\,W_2}  \,,
$$
from which we obtain solution of the dynamics
$$
  x = \frac{w_1\,\sqrt{E_1}\,\sin{(w_1 t + \phi_1)}}{1 -\, 
k_1\,\sqrt{E_1}\,\cos{(w_1 t + \phi_1)}}  \,,\qquad
  y = \frac{w_2\,\sqrt{E_2}\,\sin{(w_2 t + \phi_2)}}{1 -\, 
k_2\,\sqrt{E_2}\,\cos{(w_2 t + \phi_2)}}  \,.
$$

  We now study the superintegrability of the rational case, that is,
$w_1=n_1w_0$ and $w_2=n_2w_0$, with $n_1$ and $n_2$ positive integral numbers.
\begin{proposicion}
Let $\IK_1$ and $\IK_2$ be the following two functions
\begin{eqnarray}
  \IK_j = W_j + {\ii} n_j {w_0}\, X_j \,,\quad j=1,2\,.
{\nonumber}\end{eqnarray}
Then the complex functions  $\IK_{ij}$ defined as
$$
  \IK_{ij} = \IK_i^{n_j}\,(\IK_j^{*})^{n_i} \,,\quad i,j=1,2\,,
$$
are constants of the motion.
\end{proposicion}
{\it Proof:}
The time evolution of the functions $\IK_1$ and $\IK_2$
is given by
$$
  \frac{d}{dt}\IK_j =  \frac{d}{dt}W_j + {\ii} n_j w_0\,\frac{d}{dt}X_j
   = {\ii} n_j w_0\,\IK_j \,,\quad j=1,2\,.
$$
On the other side we have
$$
  \frac{d}{dt}\IK_{ij} = \IK_i^{n_j-1}(\IK_j^{*})^{n_i-1}
\Bigl( n_j\,\IK_j^{*}\frac{d}{dt}\IK_i +
        n_i\,\IK_i\frac{d}{dt}\IK_j^{*}  \Bigr)
$$
and from here the property follows by direct calculus.

  Thus the three functions
$$
  I_1= |\IK_1|^2\,,\quad  I_2= |\IK_2|^2  \,,{\quad}
  I_{12} = \IK_1^{n_2}\,(\IK_2^{*})^{n_1} \,,
$$
are constants of the motion.
The two first functions, $I_1$ and $I_2$, are the two one-degree of
freedom energies; concerning $I_{12}$, as it has complex value
$$
  I_{12} = I_4 + {\ii} I_3 \,,
$$
it determines not just one but two real constants of motion.
Of course, if we consider $I_3$ as the new additional constant,
$I_4$ is a function of $I_1$, $I_2$ and $I_3$.
We thus conclude that the existence of superintegrabilty and
periodic trajectories (Lissajous figures) is preserved by the nonlinearity.

  We particularize these results for the two first commensurable cases.
In the isotropic case, $w_1 = w_2 = w_0$, the two functions,
$I_3$ and $I_4$, are given by
\begin{eqnarray}
  I_3  &=& X_1 W_2 - X_2 W_1
  = \frac{(x v_y - y v_x) + (k_2 y - k_1 x) x y }
{(k_1\,v_x + k_1^2\,x^2 + w_0^2)(k_2\,v_y + k_2^2\,y^2 + w_0^2)} \,,\cr
  I_4  &=& W_1 W_2 + w_0^2 X_1 X_2
  = \frac{(v_x + k_1\,x^2)(v_y + k_2\,y^2) + w_0^2\,x y}
{(k_1\,v_x + k_1^2\,x^2 + w_0^2)(k_2\,v_y + k_2^2\,y^2 + w_0^2)} \,,
{\nonumber}\end{eqnarray}
representing the nonlinear versions of the angular momentum and
the nondiagonal component of the Fradkin tensor respectively
\cite{JaH40,Fr65}. In
fact, when  $k_1,k_2\to 0$, these two function reduce to the
appropriate expressions
\begin{eqnarray}
  \lim{}_{k\to 0}I_3   &=& \bigl(\frac{1}{w_0^4}\bigr)( x v_y - y v_x )  \,,\cr
  \lim{}_{k\to 0}I_4   &=& \bigl(\frac{1}{w_0^4}\bigr)( v_x v_y + 
w_0^2\, x y ) \,.
{\nonumber}\end{eqnarray}
Of course, for  $w_0\to 0$, we recover the $I_3$ obtained in the 
previous section;
$I_4$ just reduces to a trivial numerical constant.
\begin{eqnarray}
  \lim{}_{w\to 0}I_3  &=& \frac{(x v_y - y v_x) +  (k_2 y - k_1 x) x y }
                 {k_1 k_2 (v_x + k_1\,x^2)(v_y + k_2\,y^2)} \,,\cr
  \lim{}_{w\to 0}I_4   &=& \frac{1}{k_1 k_2} \,.
{\nonumber}\end{eqnarray}
Figure VI represents some closed trajectories in the plane $(x,y)$;
it is clear that for small energies the curves look rather similar
to the ellipses of the linear case, but for other (not so small) values of
$E$ the curves lose their elliptic shape and adopt other not so 
symmetric forms.

Now, we consider the anisotropic case  $w_1 = w_0$, $w_2 = 2 w_0$.
Then $I_3$ and $I_4$ are given by
{\openup 2pt\begin{eqnarray}
  I_3  &=& (X_1 W_2 - X_2 W_1)W_1 + w_0^2 X_1^2 X_2   \cr
  &=& \frac{(v_x + k_1 x^2)\,[\,(x v_y - y v_x) +(k_2 y - k_1 x) x 
y\,] + w_0^2 x^2 y}
{(k_1\,v_x + k_1^2\,x^2 + w_0^2)^2(k_2\,v_y + k_2^2\,y^2 + 4 w_0^2)} \,,\cr
  I_4  &=& W_1^2 W_2 + w_0^2 (4 X_2 W_1 - X_1 W_2) X_1     \cr
   &=& \frac{(v_x + k_1\,x^2)^2(v_y + k_2\,y^2) +  w_0^2\,[\,4 y v_x - 
x v_y + (4 k_1 x - k_2
y) x y\,]\, x } {(k_1\,v_x + k_1^2\,x^2 + w_0^2)^2(k_2\,v_y + 
k_2^2\,y^2 + 4 w_0^2)} \,.
{\nonumber}\end{eqnarray}}
Figure VII represents two nonlinear Lissajous figures in the plane $(x,y)$.
The situation is similar to that of figure VI;
close resemblance with the linear figures for small values of the energies
and rather strange figures for other values of $E$.
We must say that, in this case, the form is strongly dependent on the
phase difference $\phi_{12} = \phi_1 - \phi_2$.

\section{Final Comments}

  We have studied two nonlinear systems using, as starting point,
the important property of the Lagrangian origin of the second-order
Riccati equations.  In this way we could use the constant
value $E$ of the energy as an appropriate parameter for
characterizing the behaviour of these two systems.  Moreover we
have proved the existence of two--dimensional versions endowed
with superintegrability and we have obtained the explicit
expressions of the additional integrals.

  Concerning the superintegrability, we recall that most of known
superintegrable systems are superseparable systems, that is,
systems  that admit Hamilton-Jacobi separation of variables
(Schr\"o\-dinger in the quantum case) in more than one
coordinate system.
Nevertheless as all the systems studied in this paper
are nonlinear systems with a nonstandard Lagrangian
the superintegrability has been proved by considering a different approach
and without making use of the multiple separability.
In spite of this the possible separation of variables in other 
coordinate systems than Cartesian ones must be studied.

  We also mention that these two nonlinear systems can be generalized
in several different ways by making use of the Lagrangian approach.
Firstly, we recall that we have proved in Sec. 2 the Lagrangian origin
not only of the Riccati systems but also of the more general equation
(\ref{EqU(x,t)}); this means that these  more general equations could
also be studied by making use of the conservation of the energy.
Secondly, the Lagrangian (\ref{LU}) admits the following
natural generalization
\begin{equation}
  L =  \frac{1}{m(x) v_x + k\,U(x)}    \label{Lm}
\end{equation}
that leads to a nonlinear second-order equation with a
position-dependent effective mass and a additional
dissipative-looking term of the form $m'(x)v_x^2$. This new
Lagrangian (\ref{Lm}) seems interesting, not only because it
generalizes (\ref{LU}) but also because it has a more direct
geometrical interpretation since the linear function $m(x)v_x$ can
be considered as associated to the one-form $\mu=m(x) dx$.

The Lagrangian (\ref{LU}) has, as associated Hamiltonian,
the following function
\begin{equation}
  H =  - 2 \sqrt{-p_x} - k U(x)p_x   \label{HU}
\end{equation}
that, in addition to its nonnatural character (as was to be expected),
has the annoying presence of the momentum inside a root.
In the particular case of the nonlinear oscillator the Hamiltonian
is given as follows
\begin{equation}
  H = -\,\Bigl[ \frac{2 w}{k^2} \sqrt{-kp_x} + \bigl(kx^2 + 
\frac{w^2}{k}\bigr) p_x\Bigr]
  + \frac{1}{k^2} \  .
\label{Hw}\end{equation}
Nevertheless in this case we have an important property;
in fact, if we make use of the canonical transformation
$(x,p_x)\to(Q,P)$ given by
$$
  Q = \Bigl(\frac{\sqrt{2}\,}{w}\Bigr) x \sqrt{-kp_x}  \ ,\quad
  P = \Bigl(\frac{\sqrt{2\,}}{k}\Bigr) \bigl[1 - w
  \sqrt{-kp_x}\,\bigr]  \,,
$$
then we arrive at
$$
  H = \Bigl(\frac{1}{2}\Bigr)\bigl(P^2 + w^2 Q^2\bigr)   \,.
$$
Hence the very peculiar Hamiltonian (\ref{Hw}) and the
standard linear oscillator are canonically related.
Nevertheless note that this transformation has a nonpoint character and,
because of this, it cannot be directly used in the Lagrangian approach.
This brings up the question of the possible existence of similar
nonpoint transformations for Hamiltonians obtained from other
Riccati Lagrangians.
We think that this possibility is a open question to be studied.

Finally, we mention the study of the quantized versions of
all these nonlinear systems.
We note that this question must be carry out only after
the obtaining of the appropriate Hamiltonian versions
(the direct quantum study of the Lagrangian equations
appears as a difficult task).
Nevertheless as the Lagrangians are nonstandard the Hamiltonians
also appear with a unusual dependence of the momenta
(see the above expressions).
In any case the possibility or impossibility of quantizing these
systems is a matter that must be investigated.

\section*{\bf Acknowledgments.}
Support of projects  BFM-2003-02532, FPA-2003-02948,
BFM-2002-03773, and CO2-399 is acknowledged.
We are indebted to P. Leach for stimulating comments
on these nonlinear equations, and to M. Senthilvelan
for calling our attention to  some of the references.

{\small

 }

\vfill\eject
\section*{\bf Figure Captions}

\begin{itemize}

\item{}  {\sc Figure I}.{\enskip}
Plot of $x$ as a function of $t$, for $k=1$ and three different
values of the energy: $E=-0.5$, $E=-1.0$, and $E=-1.5$.

\item{}  {\sc Figure II}.{\enskip} Phase space trajectories, for
$k=1$, in the neighbourhood of the origin.

\item{}  {\sc Figure III}.{\enskip}
``Figure eight" in the plane $(x,y)$ corresponding to $k_1=k_2=1$
and energies $E_1= -1, E_2=-5$ (thick curve) and
$E_1=-1, E_2=-10$ (dash curve).

\item{}  {\sc Figure IV}.{\enskip}
Plot of $x$ as a function of $t$, for ($k=1, w=1$) and three different
values of the energy: $E=0.2$ (small thick curve), $E=0.5$ (middle curve),
and $E=0.8$ (curve with great oscillations).
The $E=0.2$ curve is very similar to a pure sine or cosine curve but, for
higher values of $E$, the plot shows clearly the effects of the nonlinearity.

\item{} {\sc Figure V}.{\enskip} Phase trajectories corresponding
to four different values of the energy ($E=0.2, 0.4, 0.6$, and
$0.8$). The trajectories are closed curves representing periodic
motions that for small values of $E$ can be considered as rather
similar to ellipses; for other values of $E$ the curves modify their
shape and they lengthen towards the upper side of the phase plane.
The motion is asymmetric in the sense that the particle moves from
right to left in a slowly way but  returns, from left to right,
with a much higher velocity that takes its maximum value at the
center point $x=0$.

\item{}  {\sc Figure VI}.{\enskip} Closed trajectories (ellipses in
the linear case) in the plane $(x,y)$ corresponding to $w_1 = 1$,
$w_2 =1$ and four different values of the energy ($E_1=E_2=0.2,
0.4, 0.6$, and $0.8$).

\item{}  {\sc Figure VII}.{\enskip} Nonlinear Lissajous figures in
the plane $(x,y)$ corresponding to ``figure eight"  trajectories
associated to $w_1 = 1$, $w_2 =2$ and energies $E_1=E_2=0.2$
(small eight-looking curve) and $E_1=E_2=0.6$ (big
butterfly-looking curve).

\end{itemize}

\vfill\eject

$$
\epsfbox{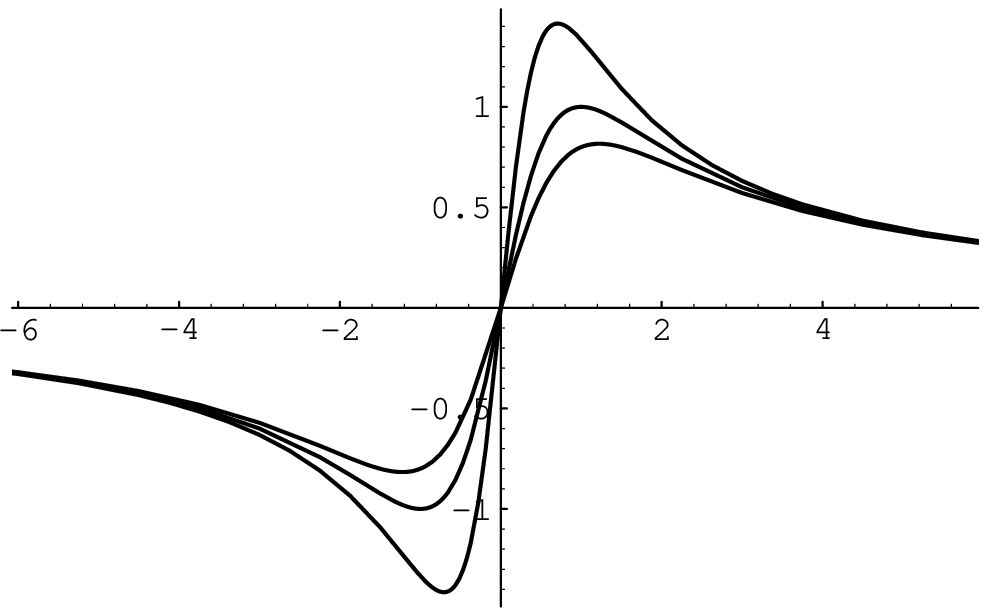}
$$

{\sc Figure I}.{\enskip} Plot of $x$ as a function of $t$, for
$k=1$ and three different values of the energy:  $E=-0.5$,
$E=-1.0$,  and $E=-1.5$.

{\vskip 40pt}
$$
\epsfbox{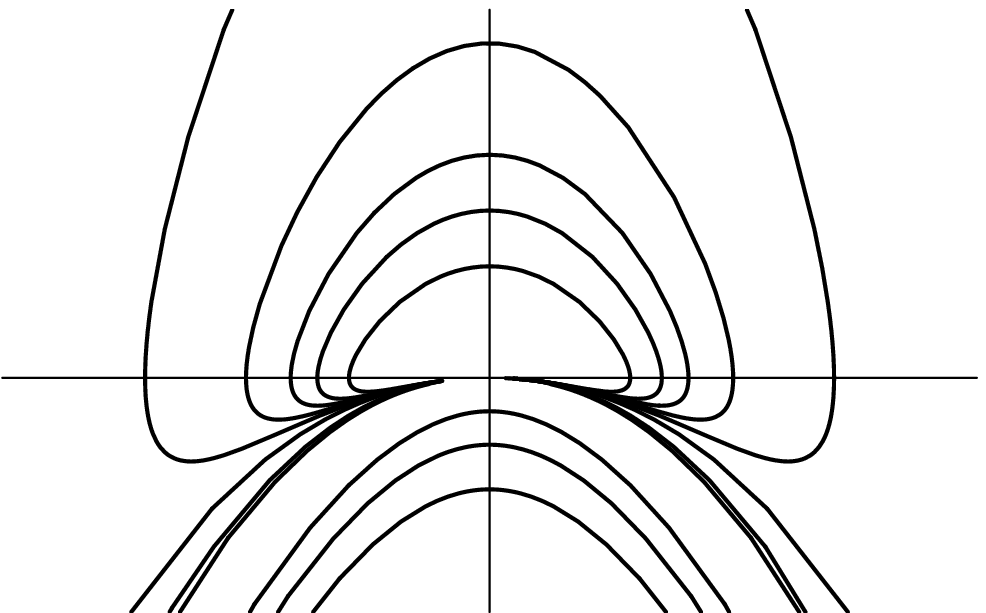}
$$

{\sc Figure II}.{\enskip} Phase space trajectories, for $k=1$, in
the neighbourhood of the origin.

\vfill\eject

$$
\epsfbox{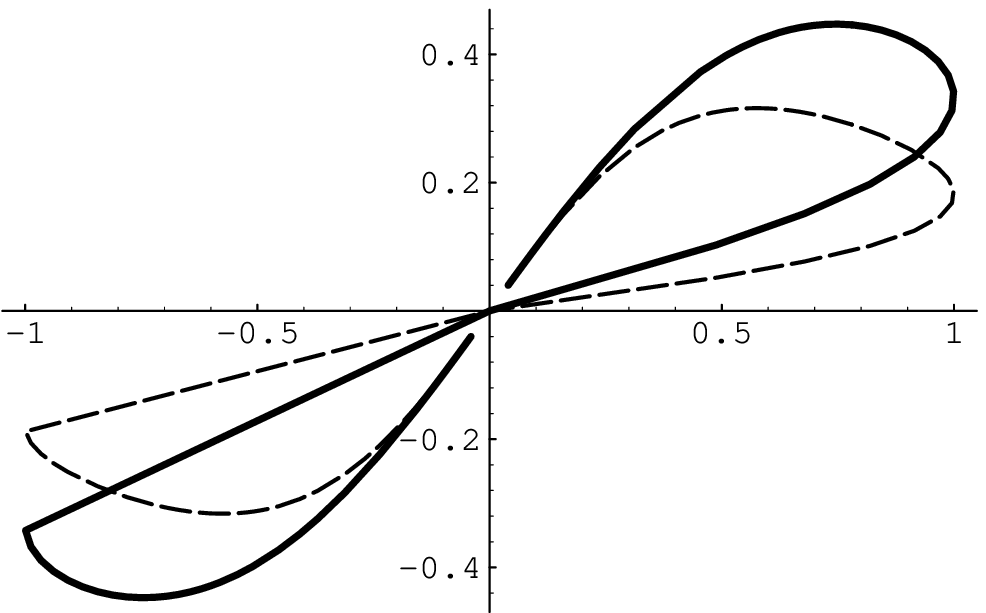}
$$

{\sc Figure III}.{\enskip} ``Figure eight" in the plane $(x,y)$
corresponding to $k_1=k_2=1$ and energies $E_1= -1, E_2=-5$ (thick
curve) and $E_1=-1, E_2=-10$ (dash curve).

{\vskip 40pt}
$$
\epsfbox{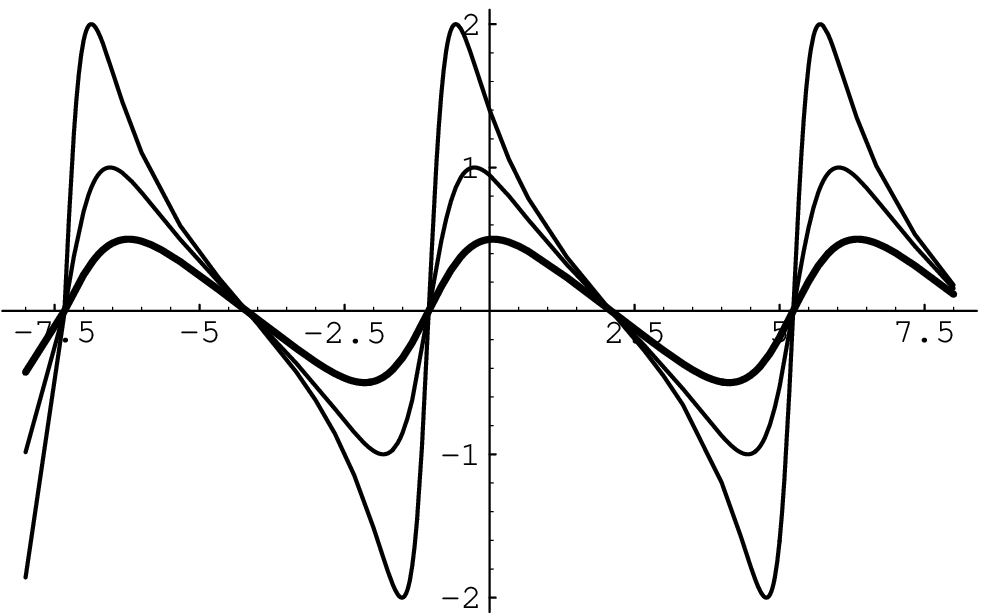}
$$

{\sc Figure IV}.{\enskip}
Plot of $x$ as a function of $t$, for ($k=1, w=1$) and three different
values of the energy: $E=0.2$ (small thick curve), $E=0.5$ (middle curve),
and $E=0.8$ (curve with great oscillations).
The $E=0.2$ curve is very similar to a pure sine or cosine curve but, for
higher values of $E$, the plot shows clearly the effects of the nonlinearity.

\vfill\eject

$$
\epsfbox{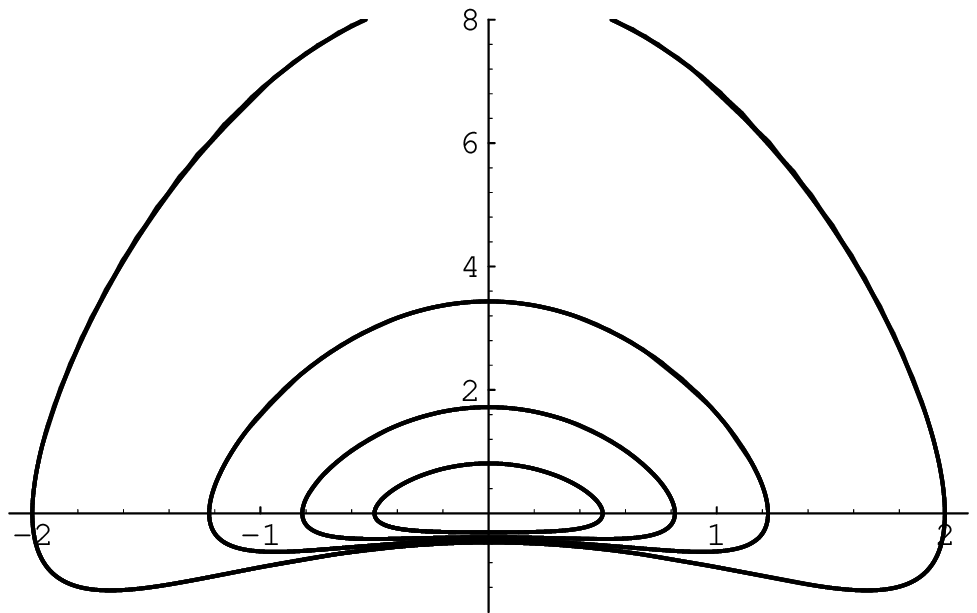}
$$

{\sc Figure V}.{\enskip} Phase trajectories corresponding to four
different values of the energy ($E=0.2, 0.4, 0.6$, and $0.8$). The
trajectories are closed curves representing periodic motions that
for small values of $E$ can be considered as rather similar to
ellipses; for other values of $E$ the curves modify their shape and
they lengthen towards the upper side of the phase plane. The
motion is asymmetric in the sense that the particle moves from
right to left in a slowly way but  returns, from left to right,
with a much higher velocity that takes its maximum value at the
center point $x=0$.

{\vskip 40pt}
$$
\epsfbox{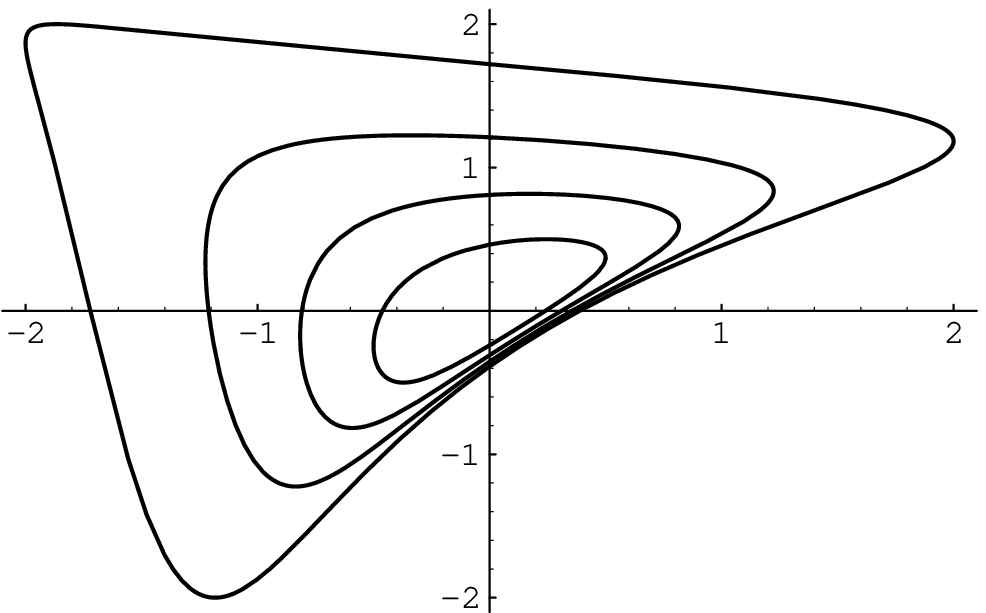}
$$

{\sc Figure VI}.{\enskip}
Closed trajectories (ellipses in the linear case) in the plane
$(x,y)$ corresponding to $w_1=1$, $w_2=1$ and four different values
of the energy ($E_1=E_2=0.2, 0.4, 0.6$, and $0.8$).

\vfill\eject

$$
\epsfbox{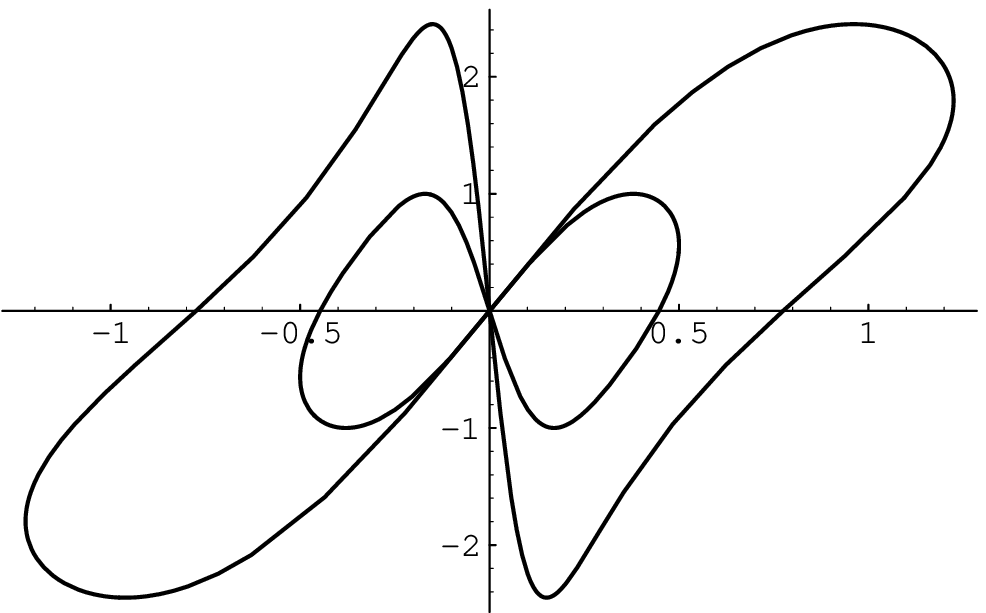}
$$

{\sc Figure VII}.{\enskip}
Nonlinear Lissajous figures in the plane $(x,y)$ corresponding to
``figure eight"  trajectories associated to $w_1 = 1$, $w_2 =2$ and energies
$E_1=E_2=0.2$ (small eight-looking curve) and $E_1=E_2=0.6$
(big butterfly-looking curve).

\end{document}